%% file: main.tex
\documentclass[10pt]{IEEEtran}
\usepackage{color}
\usepackage{wrapfig}
\usepackage{algorithm}
\usepackage{amsmath}
\usepackage{amssymb}
\usepackage{citesort}
\usepackage{fancyvrb}
\usepackage{algorithmic}
\usepackage[T1]{fontenc}
\usepackage[scaled]{helvet}
\usepackage{lscape}
\usepackage[dvips,final]{graphicx}
\usepackage{subfigure}
\graphicspath{{./}{./figures/}}
\usepackage{enumerate}
\usepackage{epstopdf}
\usepackage{float}
\usepackage{url}
\usepackage{tabularx}
\usepackage{multirow}

\input{commands.tex}

\begin{document}

\title{Peak Power Shaving for Reduced Electricity Costs in Cloud Data Centers: Opportunities and Challenges}

\author{Mehiar Dabbagh, Bechir Hamdaoui and Ammar Rayes$^{\dag}$
\thanks{An IEEE-formatted version of this article is published in IEEE Network. Personal use of this material is permitted. Permission from IEEE must be obtained for all other uses, in any current or future media, including reprinting/republishing this material for advertising or promotional purposes, creating new collective works, for resale or redistribution to servers or lists, or reuse of any copyrighted component of this work in other works.}
~\\
\small Oregon State University, \small Corvallis, OR 97331,
\small dabbaghm,hamdaoui@oregonstate.edu~\\
$^{\dag}$ \small Cisco Systems, San Jose, CA 95134,
rayes@cisco.com
}


\twocolumn
\maketitle \pagestyle{empty}
\begin{abstract}
\input{abstract.tex}
\end{abstract}
\section{Introduction}
\label{sec:PeakShaving}
\input{ElectricityBill.tex}

\subsection{Google DC Use Case: How Much Can Peak Shaving Save?}
\label{sec:Motivation}
\input{Motivation.tex}

\subsection{Why Peak Power Shaving?}
\label{sec:Why}
\input{Why.tex}

\section{Peak Power Shaving Through Energy Storage}
\label{sec:EnergyStorage}
\input{EnergyStorage.tex}

\subsection{Energy Storage Approaches}
\label{sec:Eval}
\input{Eval.tex}

\subsection{Key Challenges with Energy Storage Approaches}
\label{sec:EnergyStorage-challenges}
\input{EnergyStorage-challenges.tex}

\section{Peak Power Shaving Through Workload Modulation}
\label{sec:WorkloadModulation}
Workload modulation based peak power shaving techniques consist of modulating the DC workload in a way that reduces the peak power demand within the billing cycle~\cite{dabbagh2017shaving,wang2013data}. This can be achieved by one or a mixture of the following approaches:
\subsection{Workload Modulation Approaches}
\input{WorkloadModulation.tex}

\subsection{Key Challenges with Workload Modulation Approaches}
\label{sec:WorkloadModulation-challenges}
\input{WorkloadModulation-challenges.tex}

\section{Open Research Problems}
\label{sec:OpenProblems}
\input{OpenProblems.tex}

\section{Conclusion}
\label{sec:Conclusion}
\input{Conclusion.tex}

\section{Acknowledgement}
This work was supported in part by Cisco Systems.

\bibliography{References,References-tcc}
\bibliographystyle{IEEE}

\begin{IEEEbiography}{Mehiar Dabbagh} received the PhD degree from Oregon State University in 2016, the MS degree from the American University of Beirut in 2010 and the BS degree from the University of Aleppo in 2005, all in ECE. During his PhD studies, he interned with Cisco and with HP where he worked on developing cloud-related technologies. His research interests include: Cloud Computing, Distributed Systems, Energy Efficiency and Data Mining.
\end{IEEEbiography}

\begin{IEEEbiography}{Bechir Hamdaoui} (S'02-M'05-SM'12) is a Professor in the School of EECS at Oregon State University. He received M.S. degrees in both ECE (2002) and CS (2004), and the Ph.D. degree in ECE (2005) all from the University of Wisconsin-Madison. His research interests are in the general areas of computer networks, wireless communication, and computer security. He won several awards, including the ICC 2017 and IWCMC 2017 Best Paper Awards, the 2016 EECS Outstanding Research Award, and the 2009 NSF CAREER Award. He serves/served as an Associate Editor for several journals, including IEEE Transactions on Mobile Computing, IEEE Transactions on Wireless Communications, IEEE Network, and IEEE Transactions on Vehicular Technology. He also chaired/co-chaired many IEEE conference programs/symposia, including the 2017 INFOCOM Demo/Posters program, the 2016 IEEE GLOBECOM Mobile and Wireless Networks symposium, and many others. He served as a Distinguished Lecturer for the IEEE Communication Society for 2016 and 2017. He is a Senior Member of IEEE.
\end{IEEEbiography}

\begin{IEEEbiography}{Ammar Rayes} (S'85-M'91-SM'15) is a Distinguished Engineer / Senior Director at Cisco Services Chief Technology and Strategy Office working on the Technology Strategy.  His research interests include Network Analytics, IoT, Machine Learning and NMS/OSS.
He has authored over 100 publications in refereed journals and conferences on advances in software \& networking related technologies, 4 Books and over 30 US and International patents. He is the Founding President and board member of the International Society of Service Innovation Professionals www.issip.org, Adjunct Professor at San Jose State University, Editor-in-Chief of Advances of Internet of Things Journal, Editorial Board Member of IEEE Blockchain Newsletter, Transactions on Industrial Networks and Intelligent Systems, Journal of Electronic Research and Application and the European Alliance for Innovation - Industrial Networks and Intelligent Systems. He has served as Associate Editor of ACM Transactions on Internet Technology and Wireless Communications and Mobile Computing Journals, Guest Editor of multiple journals and over half a dozen IEEE Communication or Network Magazine issues, co-chaired the Frontiers in Service Conference and appeared as Keynote speaker at several IEEE and industry Conferences: https://sites.google.com/view/ammarrayes/home
At Cisco, Ammar is the founding chair of Cisco Services Research and Cisco Services Patent Council.  He received Cisco Chairman's Choice Award for IoT Excellent Innovation \& Execution.
He received his BS and MS Degrees in EE from the University of Illinois at Urbana and his Ph.D. degree in EE from Washington University in St. Louis, Missouri, where he received the Outstanding Graduate Student Award in Telecommunications.
\end{IEEEbiography}

\end{document}

%% file: commands.tex
\newcommand{\comment}[1]{ }



%% file: abstract.tex
An electricity bill of a data center (DC) is determined not only by how much energy the DC consumes, but especially by how the consumed energy is spread over time during the billing cycle. More specifically, these electricity costs are essentially made up of two major charges: Energy Charge, a cost based on the amount of consumed energy (in kWh), and Peak Charge, a cost based on the maximum power (in kW) requested during the billing cycle. The latter charge component is forced to encourage DCs to balance and regulate their power demands over the billing cycle, allowing the utility company to manage congestion without increasing supply.
This billing model has thus called for the development of peak power shaving approaches that reduce costs by smoothing peak power demands over the billing cycle to minimize the Peak Charge component. In this paper, we investigate peak power shaving approaches, and begin by using Google data traces to quantify and provide a real sense of how much electricity cost reduction can peak power demand shaving achieve on a Google DC cluster. We then discuss why peak power shaving is well-suited for reducing electricity costs of DCs, and describe two commonly used peak shaving approaches, namely energy storage and workload modulation. We finally identify and describe key research problems that remain unsolved and require further investigation. 

%% file: ElectricityBill.tex
When looking at the electricity bills data centers (DCs) receive from their electric utility companies at the end of each billing cycle (typically at the end of the month), we see that these bills are made up of two major charges~\cite{electricityBillModel}:
\begin{enumerate}
\item \textbf{Energy Charge.} This charge is based on the amount of energy, measured in kilo-Watt-hour (kWh), that the DC consumed within the billing cycle. In the U.S., the average energy price is 0.05 \$/kWh~\cite{electricity}. This price, of course, varies slightly from one state to another, and also changes over time within the same region.
\item \textbf{Peak Charge.}\footnote{Peak Charge is also called Demand Charge.} This charge is proportional to the maximum amount of power, measured in kilo Watt (kW), that the DC requested within the billing cycle\footnote{There exists other ways to calculate the Peak Charge that are less commonly used such as the Tiered billing model, where upon exceeding a certain power demand limit, the DC gets charged higher energy price.}. This charge represents a penalty enforced by the grid company to encourage the DC to balance the amount of power it draws over the billing cycle. The peak charge price is high and can reach up to 20 \$/kW~\cite{electricity}.
\end{enumerate}

It is clear that the DC's electricity bill is not only dependent on the amount of consumed energy, but also on how the consumed energy is being spread over time. To address this power consumption challenge, peak power shaving~\cite{even1993peak}, a well known approach for saving energy, has been exploited to reduce electricity costs of DCs, and it does so by smoothing peak power demands to minimize the Peak Charge component (e.g.,~\cite{dabbagh2017shaving}).
In this article, we focus on peak power shaving for DCs.


%% file: Motivation.tex
\begin{figure}
  \centering
  \subfigure[The power drawn from the Grid by Google DC.]{\includegraphics[width=0.8\columnwidth,height=0.35\columnwidth]{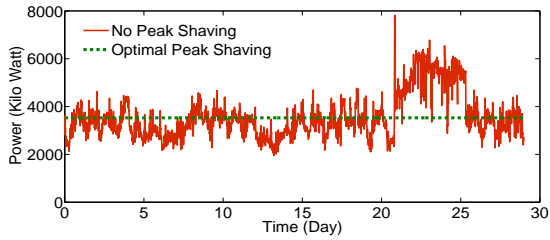}\label{fig:motivation}}\quad
  \subfigure[The electricity bill for Google DC.]{\includegraphics[width=0.8\columnwidth,height=0.35\columnwidth]{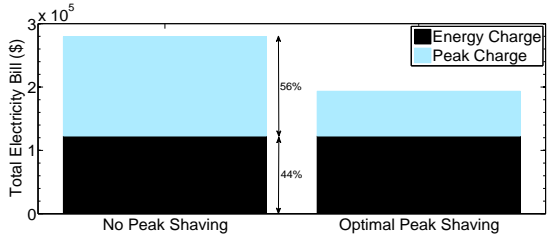}\label{fig:motivationBar}}
  \caption{No Peak Shaving versus Optimal Peak Shaving for Google DC.}
  \label{fig:GoogleMotivation}
\end{figure}
\begin{figure}
	\centering{
	\includegraphics[width=1\columnwidth,height=0.48\columnwidth]{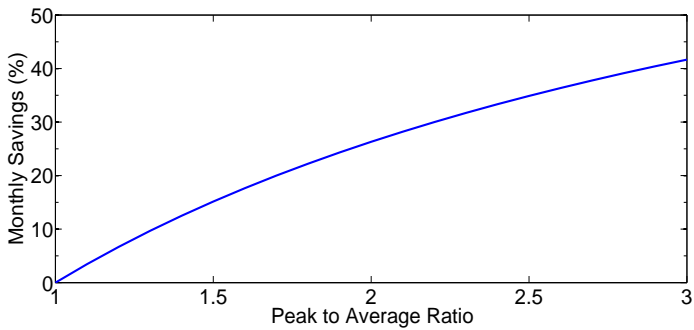}
	\caption{Monthly electricity bill savings that the optimal peak shaving case achieves w.r.t. no peak shaving case as a function of the DC's peak to average power demand ratio.}
	\label{fig:PeakToAverage}}
\end{figure}

In order to quantify the potential monetary savings that peak power shaving can achieve, we rely on real traces~\cite{Trace_Manual} collected from a Google DC made up of around 12.5K servers. The traces report the amount of computing resources (e.g. CPU, memory) that were used by the different tasks/jobs submitted to a Google DC over a 29-day period. For each point in time, we track the aggregate amount of computing resources used by the DC's tasks and estimate the minimum number of servers needed to host those tasks. The power demanded by each ON server is then estimated based on the server's CPU utilization, whereas the servers that are not hosting any tasks are assumed to be switched off to save energy and thus don't consume any power. In order to account for the power consumed by the non-computing infrastructure, the DC is assumed to have a Power Usage Efficiency (PUE) of 1.7, which means that for every 1 Watt that the DC consumes on IT-work, an additional 0.7 Watt is consumed by the non-computing infrastructure (e.g. cooling devices, facility lighting, etc.). Based on these calculations, Fig. \ref{fig:motivation} plots the total power drawn by Google DC over time, which is referred to as the 'No Peak Shaving' case. The figure also plots the power demands for the 'Optimal Peak Shaving' case, where the amount of energy the DC demanded during the entire trace period was spread evenly over the entire billing cycle, thus yielding the minimal/optimal Peak Charge. Next, we rely on real electricity price~\cite{electricity} and show in Fig. \ref{fig:motivationBar} the total electricity bill corresponding to each of the two cases ('No Peak Shaving' and 'Optimal Peak Shaving'), with the bill being made up again of the Energy Charge and the Peak Charge. Observe the high contribution (56\%) of the Peak Charge to the total electricity bill for the No Peak Shaving case. By minimizing the Peak Charge, the Optimal Peak Shaving case can reduce the total electricity bill by 31\%, which translates into savings of \$86K per month. This study clearly shows that peak shaving has a great potential for achieving significant monetary savings. Of course these monthly monetary savings depend on the peak to average ratio of the power demanded by the DC within the billing cycle. In order not to limit ourselves only to Google traces, we plot in Fig. \ref{fig:PeakToAverage} the monetary savings that the optimal peak shaving case achieves with respect to the no peak shaving case as a function of the peak to average power demand ratio. The figure clearly shows that the peak shaving monetary savings increase quickly as this ratio increases.

%% file: Why.tex
\begin{figure}
	\centering{
	\includegraphics[width=1\columnwidth,height=0.48\columnwidth]{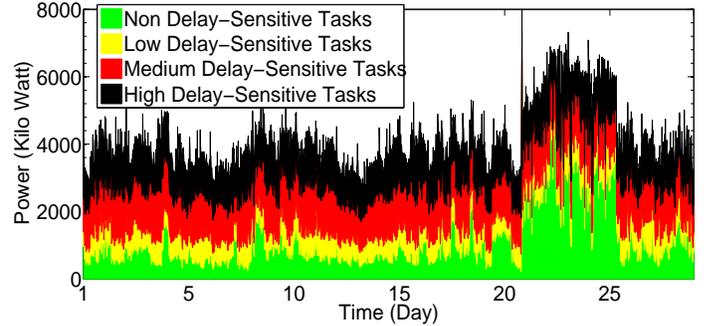}
	\caption{Breakdown of the power consumed by the different workload classes for Google DC.}
	\label{fig:WorkloadBreakdown}}
\end{figure}

Several incentives and reasons we list next make DCs appropriate for and beneficial from adopting peak power shaving:
\begin{enumerate}
\item \textbf{High Energy Consumption.} Large IT companies such as Google, Amazon and Facebook have multiple DCs distributed around the world, with each DC containing tens of thousands of servers. Any time you are watching a Netflix movie, doing a Google search, or browsing photos on Facebook, some computations are being executed on one of those DCs' servers, and this server is consuming energy. It is thus no surprise that DCs consume enormous amounts of energy, incurring high electricity costs. Google has revealed that its DCs around the globe continuously draw 260 million Watts, an amount that is enough to power 200K households~\cite{Google-stat11}. This electricity consumption costed Google \$5 billion in the 2\textsuperscript{nd} quarter of 2014~\cite{Google-stat11}. The energy consumption of U.S. DCs in 2013 is estimated to 91 billion KWH, an amount that is twice the electricity needed to power the whole city of New York~\cite{DC-stat14}. DCs' energy consumption is growing rapidly (by 10-12\% per year) with the increasing popularity of services and applications that rely on DCs. These are strong financial incentives that alert DC owners to find effective ways to cut down on their electricity costs.

\item \textbf{Supporting Infrastructure.} Modern DCs are highly automated and are supplied by monitoring infrastructure that provides logs of the power and resource utilization at the task, server and DC levels. These DCs' monitoring and automation capabilities allow to apply control techniques for peak shaving that are responsive and reactive.
    Furthermore, DCs are (typically) equipped with Diesel generators and energy storage devices, aka Uninterruptible Power Supply (UPS), that  serve as backup sources of power during power outages.
    In addition, DCs are usually distributed across different geographical locations and have the ability to migrate active tasks from one server to another, within the same or across different DC, with minimal service disruption. Finally, DC servers can operate at varying power levels, with each level having power and performance tradeoffs.
    All of these DC capabilities (UPS, migration, tunable operational power) ease the adoption of peak sharing techniques, and thus provide great incentives for applying them.

\item \textbf{Workload Heterogeneity and Flexibility.} DCs' workloads are made up of tasks with different service requirements.
    The Google traces~\cite{Trace_Manual}, for instance, classifies tasks into four classes, based on their delay sensitivity: no-, low-, medium- and high-delay sensitive. Figure~\ref{fig:WorkloadBreakdown} shows a breakdown of the power consumed by the tasks belonging to these four classes for the Google DC. This workload heterogeneity and delay flexibility open up great doors for using peak shaving as an approach for reducing electricity costs in DCs.
\end{enumerate}

The rest of the article is organized as follows. We begin, in Section \ref{sec:Motivation}, by using real data traces~\cite{Trace_Manual} provided by Google to quantify and assess how much money peak shaving can save for one of Google's DC.
The article then discusses and argues, in Section \ref{sec:Why}, how peak power shaving is well-suited for DCs.
The two main peak shaving techniques, energy storage and workload modulation, are explained in Sections \ref{sec:EnergyStorage} and \ref{sec:WorkloadModulation}, respectively.
The article finally identifies and describes, in Section \ref{sec:OpenProblems}, key problems that remain unsolved, and provides some concluding remarks in Section \ref{sec:Conclusion}.

%% file: EnergyStorage.tex
Energy storage-based peak shaving techniques store energy during low power demand periods, and use it later to power the DC (partially or fully) during peak power demand periods~\cite{dabbagh2016peak}. What makes these techniques attractive is the fact that they do not result in any performance degradation of the active DC tasks. These techniques, however, require infrastructure to store energy, but the good news is that DCs are already equipped with UPS devices,
which although are initially included to serve as backup when power goes down, they can also be used for peak shaving. The use of these UPS devices for peak shaving does not contradict with their fault tolerance role, as they usually can store enough energy to power the DC at full capacity for half an hour during power outage; the transition time, the time it takes a diesel generator to start pumping power to the DC, is only about few tens of seconds~\cite{dieselUPS}.
Thus, UPS devices can always store enough energy to power the DC during power outages while also using the remaining capacity for peak power shaving purposes.

\subsection{UPS Topologies}
UPS topologies can be {\em centralized} or {\em distributed}.
In distributed topologies, one small UPS device is mounted either on each server, as it is the case for Google's modern DCs---this is called \textit{server-level} distributed topology---or on each rack of few servers as it is the case for Facebook's DCs---this is called \textit{rack-level} distributed topology. In centralized topologies, a large, shared UPS device provides power to the entire DC.

\begin{table}
\centering
\caption{Centralized vs. distributed UPS topology.}
\label{table:UPS_Comparison}
\begin{tabular}{ | c | c | c |}
 \hline
  & Centralized Topology & Distributed Topology \\
 \hline\hline
 Reliability & Less Reliable & More Reliable \\
 \hline
 Single-point-of-failure & Yes & No \\
 \hline
 Scalability & No & Yes \\
 \hline
 Conversion Losses & Higher & Lower \\
 \hline
 Locked-in Energy & No & Yes \\
 \hline
 Management Complexity & Lower & Higher \\
 \hline
 Battery Lifetime & Longer & Shorter \\
 \hline	
\end{tabular}
\end{table}

\subsubsection{Centralized vs. Distributed UPS Topologies}
We discuss next the pros and cons of these UPS topologies (summarized in Table \ref{table:UPS_Comparison}):
\begin{itemize}
\item {\bf Reliability and fault tolerance:} Compared to centralized topologies, distributed topologies do not suffer from single point-of-failure, and require less wiring as UPS devices are mounted closer to the servers, making them more reliable and fault tolerant topologies.

\item {\bf Conversion Losses:} Centralized topologies yield conversion losses, because power drawn from the grid gets converted from AC-to-DC first in order to be stored in battery and then converted back to AC from DC to power the servers. Each of these two conversions yields some power loss. Distributed topologies, whether server-level or rack-level, on the other hand, do not involve conversion, as the battery feeds DC power directly to the servers.

\item {\bf Locked-in Energy:} In distributed topologies, each battery can supply power only to the server or the rack of servers it is connected to, limiting thus the amount of power that can be shaved in the entire DC.
This is because there could be a scenario where some lightly loaded servers (with low energy needs) have large amounts of energy stored in their local UPS devices, but locked as there is no way to route it to other highly loaded servers in the DC that might need energy to shave their demands.


\item {\bf Management Complexity:} The distributed topology is harder to manage as it requires implementing a monitoring system that has a consistent view of multiple UPS devices in the DC in addition to the need to develop efficient control policies that decide what UPS device to charge/discharge.
\item {\bf Capital Expenses and Resource Provisioning:} The distributed topology is more attractive for growing companies as they only need to buy a small UPS for each server in their DC where additional small UPS devices can be bought later when additional servers are added to the DC. This is usually preferred over the high upfront payment associated with buying a single centralized UPS device with enough capacity to support the DC's predicted growing demands.

\item {\bf Lifetime:} Normally, distributed batteries have a shorter lifetime and need to be replaced more often, as they operate in a high-temperature environment since they are surrounded by thousands of servers, whereas the centralized UPS device is usually located in a separate, cooler room.
\end{itemize}

\subsubsection{Server-level vs. Rack-level Distributed UPS Topologies}
Server-level distributed topologies yield lesser losses than rack-level ones, as in the former case, the battery sits right after the server's power supply unit, thus minimizing wiring losses. The rack-level case, however, reduces the amount of locked-in energy as the stored energy can be supplied to any server in the rack rather than just being exclusively supplied to only a single server, as in the server-level case.


%% file: Eval.tex
Different control strategies have been proposed to decide when and how much energy to charge/discharge into the UPS batteries. The following summarizes the main strategies, including approaches introduced by the authors of this article:
\begin{itemize}
\item {\bf Threshold Strategy:} This technique~\cite{thresh3} operates the DC at a power level below the peak by periodically comparing the DC's power demand with a specified threshold. If the demand is below the threshold, then the difference (capped by the battery's maximal capacity and charging rate) is charged into the battery within that period. Otherwise, the battery discharges the difference (capped by the battery's maximum discharge rate and the maximum amount of stored energy) within that period. This approach is widely adopted in industry due to its simplicity. Its main limitation is that it is highly sensitive to the selected threshold.
\item {\bf Predictive Strategy:} This technique~\cite{dabbagh2017shaving} predicts future power demand of DC using recent demand history, and uses these predictions to decide how much energy to charge/discharge. These decisions are made by formulating the problem as an optimization problem whose objective is to minimize the DC's electricity bill, while accounting for leakage and conversion losses and meeting the capacity, charging rate, and discharging rate constraints.
\end{itemize}

\begin{figure}
	\centering{
	\includegraphics[width=1\columnwidth,height=0.48\columnwidth]{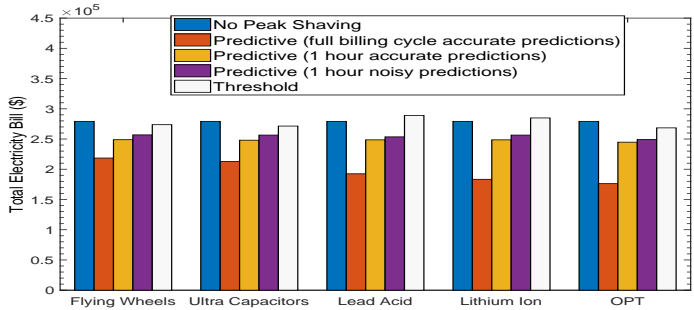}
	\caption{Comparison of Existing Energy Storage Control Strategies based on Google traces for different battery types.}
	\label{fig:comparison}}
\end{figure}

Fig. \ref{fig:comparison} shows the total electricity bill of Google DC~\cite{Trace_Manual} when operating without peak shaving, with the Threshold Strategy, and with the Predictive Strategy for different types of batteries: Lead-Acid, Lithium-Ion, Flying Wheels and Ultra-Capacitors batteries (see \cite{dabbagh2017shaving} for further details on the specs of considered batteries).
Our evaluations of the monetary bill shown in Fig. \ref{fig:comparison} are based on (i) 29-day worth of real data traces collected from a Google DC cluster, (ii) real power prices given in~\cite{electricity}, and (iii) energy consumption calculation models provided in~\cite{electricityBillModel}, where the prices used for calculating the Energy Charge and the Peak Charge are 0.05 \$/kWh and 20 \$/kW respectively, and the Peak Charge is calculated based on dividing the billing cycle into slots each of length $15$ minutes. More details can be found in \cite{dabbagh2017shaving}.
Since the behavior of the Predictive Strategy depends on the length of the prediction window and on the accuracy of the predictions, we evaluate the Predictive Strategy while considering: a) noisy predictions of the power demands in the coming hour, b) perfectly accurate predictions of the power demands in the next hour, and c) perfectly accurate predictions of the power demands for the whole billing cycle (this represents the optimal upper bound). Observe from Fig. \ref{fig:comparison} that predicting the power demands for a short duration of one hour saves a significant amount of electricity cost as compared to No Peak Shaving and these savings are higher than the Threshold technique and are very close to the optimal case. Obviously the more accurate the predictions and/or the longer the length of the prediction window, the greater the savings that the Predictive Strategy can achieve.

%% file: EnergyStorage-challenges.tex

\begin{itemize}
\item {\bf Energy Loss:} Some of the stored energy gets lost during AC-to-DC and DC-to-AC conversion stages. In addition, stored energy decays over time due to energy leakage or battery self-discharge, and the loss varies depending on the battery type, the number of charging/discharging cycles, and the temperature. These losses should then be taken into account when making peak shaving design decisions.

\item {\bf Workload Uncertainty:} DCs' power demands vary over time, and hence, effective peak power shaving would require prior knowledge (e.g. via prediction) of when low and high power demand periods will occur.

\item {\bf Limited Battery Capacity:} Batteries have limited energy storage capacities and their charging/discharging rates are bounded. These limitations should be accounted for when designing peak shaving strategies, as they impact significantly how such strategies perform.

\item {\bf Locked-in Energy:} This poses a challenge only for distributed UPS topologies, which require control policies that minimize the amount of locked energy so as to enhance peak shaving performance.

\item {\bf Battery Lifetime:} Charging/discharging UPS devices increases energy losses and shorten batteries' lifetimes, requiring frequent battery replacement. These replacement costs should be carefully accounted for when designing peak shaving approaches.
\end{itemize}


%% file: WorkloadModulation.tex
\begin{itemize}
\item {\bf Workload Dropping:} This approach drops the workload during peak periods to reduce the Peak Charge. This is done by either rejecting low-priority tasks or stopping some already running tasks. Clearly, this may cause service disruption of submitted tasks and is only efficient when the reduction in the Peak Charge outweighs the revenue lost from dropping such tasks. However, recent advancements in virtualization technology allow to checkpoint running tasks before dropping them, so that to resume from that checkpoint when tasks are launched again. There is, however, an incurred overhead in terms of time, energy and storage that is spent to perform this checkpoint-resume operation.

\item {\bf Request Redirection:} It is very common to have multiple instances of the same application running across multiple DCs. This technique basically exploits this geo diversity and selects which DC should handle the application request such that the Peak Charge across all DCs is reduced. One way for redirecting requests is to have a load balancer that receives all the requests and selects which application instance should handle each received request. Another way to redirect requests is to update the Domain Name Service (DNS) records of the application, which informs the clients to which instance their request should be routed to based on the IP address found in the DNS record.

\item {\bf Dynamic Voltage and Frequency Scaling (DVFS):} This approach adjusts operating speed and voltage of some of the server's components (e.g. CPU) to reduce power consumption, but at the cost of incurring some delays to all the tasks hosted on the server accessing that component. DVFS is supported by most of DC's servers where the server's components can normally operate at different voltage and frequency levels.

\item {\bf Resource Scaling:} This approach reduces the amount of allocated resources for some of the running tasks during peak periods, leading to power consumption reduction, but at the cost of some performance degradation. There are two types of resource scaling: $a)$ Horizontal, where the number of application's instances is reduced. An example of horizontal scaling would be to reduce (scale out) the number of web servers from 8 to 5 instances. $b)$ Vertical, where the amount of allocated resources per application's instance is reduced. An example of vertical scaling would be to reduce (scale down) the allocated CPU resources of a web server instance from 2 cores to 1.

\item {\bf Inter-DC Workload Migration:} This approach migrates tasks from a cluster that is experiencing peak power demands to another lightly loaded cluster, often located in a different geographical location.
\end{itemize}

%% file: WorkloadModulation-challenges.tex
\begin{itemize}
\item {\bf Task Selection:} DCs host a massive number of tasks, typically in the order of millions and varies depending on the size of the DC. One challenge would be how to decide which task is to be delayed or dropped. Several factors, including task priority, Service-Level Agreements (SLAs), generated revenue, fairness, can affect such decisions.

\item {\bf Variable Delays:} Reducing the allocated resources or the operating voltage/frequency varies from a workload to another.
    For example, reducing CPU's operating frequency/voltage might cause insignificant performance degradation for memory-intensive tasks, but high delays for CPU-intensive tasks. This delay variability makes it very challenging to maintain SLAs at their required levels.

\item {\bf Application Constraints:} Applications are typically made up of multiple micro services (multiple tasks) that interact with each other, and that each may have some scheduling constraints. Example of these constraints include placing different tasks on the same server, or on servers belonging to the same rack to guarantee low networking delay. The constraint could also be not to place some tasks (e.g., multiple replicas of the applications) on the same server, for instance, for fault tolerance purposes.
    It is therefore important to ensure that these constraints are met when making migration decisions.

\item {\bf Workload Uncertainty:} The variability of DCs' workload demands requires careful attention when deciding when the workload should be delayed. This is to avoid shifting the workload from a period perceived to have a high demand to another perceived to experience low demands but turns out to actually have higher demands than the period from which the demand is shifted. When this happens, it will increase the Peak Charge in addition to causing performance delays when compared to not shifting workload at all.
\end{itemize} 

%% file: OpenProblems.tex
The following are key open problems, related to DC peak power shaving, that require further research investigation:

\paragraph{Workload Modeling} Workload modeling remains an open area that has not received much attention. Questions, such as how much delay does it incur if and when the resources allocated to a task are to be reduced, and how much power does it save and how much delay does it incur if and when the operating frequency/voltage of a server is reduced by a certain amount, have not been answered yet.
Answering these questions requires selecting an appropriate benchmark of cloud applications and studying the effects of varying the allocated resources or the server's operating frequency/voltage under varying server platforms.
Workload modeling and benchmarking are essential to developing workload modulation techniques that can provide effective ways of shaving the power demand peaks.

\paragraph{Fairness} When deciding which tasks should be delayed or dropped, it is important to consider strategies that ensure fairness among the tasks handled by the DC. This requires defining and introducing appropriate fairness metrics, as well as developing methods that can quantify and enforce them.

\paragraph{Workload Prediction and Resource Management} Efficient peak shaving strategies should rely on machine learning (e.g., deep reinforcement learning) to provide accurate predictions of a DC's future power demands~\cite{wang2019smart}. It is worth mentioning that many off-the-shelf prediction techniques, such as those that rely on weighted averages of recent demands to predict future demands, might not work because weighted averages are always guaranteed not to be greater than the maximum of previously observed demands. Hence, such techniques would never predict a peak higher than the demand seen so far, preventing the predicted workload from not being high enough for the controller to be worth shaving.
Also, many prior works proposed for reducing the DCs' energy ignore peak shaving.
For example, prior scheduling algorithms (e.g., bin-packing based) aim to reduce the number of running servers~\cite{dabbagh2015energy,dabbagh2014energy}, but completely ignore the energy stored in the UPS batteries attached to those servers. This limits the amount of power that can be shaved and thus the electricity cost that can be saved.

\paragraph{Peak Shaving Technique Selection} We have not seen in the literature many frameworks that apply joint workload modulation and energy storage control approaches for peak shaving. These two techniques are not contradictory but rather complimentary, as it is possible during a high demand period to delay some of the tasks and also to rely on a battery to partially supply the demanded power. The problem boils down to how to select when to apply each technique. Also, there exists many approaches that can be used to modulate the workload for peak shaving, each incurring its own cost, and hence, there is a need to develop a controller that decides which approach within workload modulation should be applied and to which tasks.

\paragraph{Workload Revenue Quantification} This is important for public cloud DCs, where deciding which tasks to drop depends on how long the tasks run; that is, it would be more beneficial, from a revenue viewpoint, to drop short-running tasks instead of long-running ones.
However, clients do not specify in advance how long their tasks are going to run for. Additionally, if a client's task gets dropped often, then the client might be encouraged to switch to another public cloud provider. This requires developing smart techniques that select which tasks to drop with the objective of maximizing the long-term revenue.

\paragraph{Hybrid DC Architectures} The emergence of edge cloud offloading warrants future DCs to evolve towards hybrid architectures that combine conventional DCs with edge clouds, where edge servers will be leveraged to mitigate the resource limitation challenges of emerging IoT devices and the stringent latency requirements of newly emerging wireless applications~\cite{guo2018mobile}. Not much research has been devoted to studying energy-aware scheduling techniques that are suited for such hybrid DC architectures.


%% file: Conclusion.tex
In this paper, we use Google data traces to investigate and quantify the major charge components, Energy Charge and Peak Charge, that make up an electricity bill of a Google DC. Our findings show that the Peak Charge component contributes to about 56\% of the DC's total electricity costs, demonstrating the importance of adopting peak demand shaving techniques in DCs as a potential approach to reduce the DCs' electricity costs.
We then describe two commonly used peak shaving approaches, namely energy storage and workload modulation, as well as the challenges these approaches face. We finally identify and discuss key open research problems and challenges, pertaining to peak demand shaving in DCs, that require further research investigation and attention. 